\documentclass[11pt]{article}
\usepackage{hyperref}
\usepackage{amsmath}
\usepackage{graphicx,epsfig}
\usepackage{subfigure}
\usepackage{color}
\usepackage{bm}
\usepackage{multirow}
\usepackage{verbatim}
\usepackage{amssymb}
\usepackage{amsfonts}
\usepackage{amsmath}

\usepackage{authblk}

\usepackage{enumitem}

\usepackage[displaymath]{lineno}

\usepackage{multirow}

\usepackage{graphics}
\usepackage{epsfig}
\usepackage{pgf}
\usepackage{pgfpages}
\usepackage{xcolor}
\usepackage{pifont}

\graphicspath{{./Figs/}}

\usepackage{float}     

\usepackage{titletoc}
\usepackage[raggedright,pagestyles]{titlesec}

\usepackage{fancyhdr}
\usepackage{lastpage}
\usepackage{layout}

\usepackage[footnotesep=10mm]{geometry}

\usepackage{caption2}

\usepackage{setspace}

\makeatletter
\renewcommand\@biblabel[1]{\footnotesize$[{#1}]$\selectfont}
\makeatother  %

\titlecontents{part}[0em]{\vskip3mm \bf B.\contentslabel{9em}}{}
{\bf \hspace*{0em}}
{\titlerule*[0.8pc]{.}\contentspage}[\addvspace{0.5ex}]

\titlecontents{section}[1em]{\it \contentslabel{0em}}{\hspace*{1.5em}}
{\it\hspace*{2.3em}}
{\titlerule*[0.8pc]{.}\contentspage}[\addvspace{0.5ex}]

\titlecontents{subsection}[2.7em]{\contentslabel{0em}}{\hspace*{2.5em}}{\hspace*{2.3em}}
{\titlerule*[0.8pc]{.}\contentspage}[\addvspace{0.5ex}]

\titleformat{\part}{\raggedright\large\bfseries}{\bf\normalsize{B.\thepart} }{1em}{}
\titleformat{\section}{\raggedright\normalsize\bfseries}{\bf\normalsize{\S\,\thesection}.}{1em}{}
\titleformat{\subsection}{\raggedright\normalsize\bfseries}{\thesubsection}{1.5em}{}
\titleformat{\subsubsection}{\raggedright\normalsize\itshape}{\thesubsubsection}{1em}{}

\usepackage{hyperref}

\linethickness{0.6pt}

\newcommand\PICline{\setlength{\unitlength}{2.5mm}
\begin{picture}(2,1)
\put(0.9,-0.5){\line(0,1){2}}
\end{picture}}

\newcommand\PICcircle{\setlength{\unitlength}{3mm}
\begin{picture}(1.8,1)(0.2,0.5) \put(1,1){\circle{1.2}}
\end{picture}}

\newcommand\PIClinecircle{\setlength{\unitlength}{2.5mm}
\begin{picture}(3.3,1.2)(0.5,0.5)
\put(1,0){\line(0,1){2}}
\put(2.6,1){\circle{1.2}}
\end{picture}}

\newcommand\PICunorientLRsplit{\setlength{\unitlength}{0.6mm}
\begin{picture}(9,6)(-1,2)
\qbezier(0,0)(4,4)(0,8) \qbezier(6,0)(2,4)(6,8)
\end{picture}}

\newcommand\PICunorientUDsplit{\setlength{\unitlength}{0.6mm}
\begin{picture}(11,6)(-1,1)
\qbezier(0,6)(4,2)(8,6) \qbezier(0,0)(4,4)(8,0)
\end{picture}}

\newcommand\PICunorientpluscross{\setlength{\unitlength}{2mm}
\begin{picture}(2.9,1.9)(-1.3,-0.6)
\qbezier(-1,-1)(0.0,0.0)(1,1)
\qbezier(-0.3,0.3)(-0.8,0.8)(-1.0,1.0)
\qbezier(0.3,-0.3)(0.8,-0.8)(1.0,-1.0)
\end{picture}}

\newcommand\PICunorientminuscross{\setlength{\unitlength}{2mm}
\begin{picture}(2.9,1.9)(-1.3,-0.6)
\qbezier(1,-1)(0.0,0.0)(-1,1)
\qbezier(0.3,0.3)(0.8,0.8)(1.0,1.0)
\qbezier(-0.3,-0.3)(-0.8,-0.8)(-1.0,-1.0)
\end{picture}}

\newcommand\PICopengammaplus{\setlength{\unitlength}{0.5mm}
\begin{picture}(11,8)(-2,-6)
\qbezier(1.4,-5.2)(2.0,-6.2)(3.2,-6.1)
\qbezier(3.2,-6.2)(6.0,-6.2)(6.2,-3.1)
\qbezier(6.2,-3.1)(5.9,-0.3)(3.1,-0)
\qbezier(3.1,-0)(0.3,-0.3)(0.0,-2.9)
\put(0.0,-3.1){\line(0,-1){7}} \put(0,0.8){\line(0,1){4}}
\end{picture}}

\newcommand\PICopengammaminus{\setlength{\unitlength}{0.5mm}
\begin{picture}(11,8)(-2,0)
\qbezier(1.4,5.2)(2.0,6.2)(3.2,6.1)
\qbezier(3.2,6.2)(6.0,6.2)(6.2,3.1)
\qbezier(6.2,3.1)(5.9,0.3)(3.1,0)
\qbezier(3.1,0)(0.3,0.3)(0.0,2.9)
\put(0.0,3.1){\line(0,1){7}} \put(0,-0.8){\line(0,-1){4}}
\end{picture}}

\newcommand\PICclosegammaplus{\setlength{\unitlength}{0.5mm}
\begin{picture}(16,8)(-8,-2)
\qbezier(1.4,-1.8)(2.0,-2.7)(3.2,-2.9)
\qbezier(3.2,-2.9)(6.0,-2.7)(6.2,-0.2)
\qbezier(6.2,-0.2)(5.9,2.6)(3.1,2.9)
\qbezier(3.1,2.9)(0.3,2.6)(0.0,0.0)
\qbezier(-1.4,1.8)(-2.0,2.7)(-3.2,2.9)
\qbezier(-3.2,2.9)(-6.0,2.7)(-6.2,0.2)
\qbezier(-6.2,0.2)(-5.9,-2.6)(-3.1,-2.9)
\qbezier(-3.1,-2.9)(-0.3,-2.6)(0.0,0.0)
\end{picture}}

\newcommand\PICclosegammaminus{\setlength{\unitlength}{0.5mm}
\begin{picture}(16,8)(-8,-2)
\qbezier(-1.4,-1.8)(-2.0,-2.7)(-3.2,-2.9)
\qbezier(-3.2,-2.9)(-6.0,-2.7)(-6.2,-0.2)
\qbezier(-6.2,-0.2)(-5.9,2.6)(-3.1,2.9)
\qbezier(-3.1,2.9)(-0.3,2.6)(0.0,0.0)
\qbezier(1.4,1.8)(2.0,2.7)(3.2,2.9)
\qbezier(3.2,2.9)(6.0,2.7)(6.2,0.2)
\qbezier(6.2,0.2)(5.9,-2.6)(3.1,-2.9)
\qbezier(3.1,-2.9)(0.3,-2.6)(0.0,0.0)
\end{picture}}

\begin{document}
\title{Tackling tangledness of cosmic strings by knot polynomial topological invariants}
\author[1]{\small Xinfei LI}
\author[2,1]{\small Xin LIU \thanks{Corresponding author: xin.liu@bjut.edu.cn}}
\author[1]{\small Yong-Chang HUANG}
\affil[1]{\small Institute of Theoretical Physics, Beijing University of Technology,  Beijing 100124, China}
\affil[2]{\small Beijing-Dublin International College, Beijing University of Technology,  Beijing 100124, China}
\date{\today}
\maketitle
\begin{abstract}

\setlength{\parindent}{0pt} \setlength{\parskip}{1.5ex plus 0.5ex
minus 0.2ex} 
Cosmic strings in the early universe have received revived interest in recent years. In this paper we derive these structures as topological defects from singular distributions of the quintessence field of dark energy. Our emphasis is placed on the topological charge of tangled cosmic strings, which originates from the Hopf mapping and is a Chern-Simons action possessing strong inherent tie to knot topology. It is shown that the Kauffman bracket knot polynomial can be constructed in terms of this charge for un-oriented knotted strings, serving as a topological invariant much stronger than the traditional Gauss linking numbers in characterizing string topology. Especially, we introduce a mathematical approach of breaking-reconnection which provides a promising candidate for studying physical reconnection processes within the complexity-reducing cascades of tangled cosmic strings.
\end{abstract}


\vspace*{10mm}

\section{Introduction}

Cosmic strings were first proposed by Kibble in 1976 from the field theoretical point of view \cite{Kibble:1976}. As one-dimensional topological defects with zero width, their formation took place through a symmetry breaking phase transition (the Kibble mechanism) of an abelian Higgs model in the early universe, at the quenching stage after the cosmological inflation \cite{Rajantie:2003wi}.

String theory redefines the significance of cosmic strings. A first prediction was based on F-strings, stating that a string could be produced in the early universe and stretched to macroscopic scale. But that mechanism has two potential unpleasant prospects \cite{Witten:1985}. One is, as asserted by type-I and heterotic superstring theories,  that these F-strings would be produced as disintegrated small strings and appear as boundaries of domain walls, which would therefore inevitably collapse before growing into cosmic scale, due to their high tension. The other is that, in the Planck energy scale, the strings would be created prior to the cosmological inflation, and therefore get diluted away and unobservable in consequence during the inflation.

Developments of string theory overcame the above difficulties and strengthened the linkage between cosmic string and superstring theories. New one-dimensional strings were discovered, including the D-strings which correspond to F-strings via the S-duality, as well as the NS- and M-branes which have only one non-compact dimension (--- the other dimensions being wrapped into an internal compact Calabi-Yau manifold with throats \cite{Kachru:2003sx}). Low-tension strings which are stable in expanding universe are achievable now, since large compact dimensions and warp factors would suppress string tension to an observable energy scale. In 2002 Tye predicted an scenario of brane inflation where low dimensional D-branes with one non-compact dimension is produced \cite{Sarangi:2002yt}. Later, Polchinski suggested that a string could be stretched to intergalactic scale in expanding universe \cite{Copeland:2003bj,Polchinski:2004hb}. As Kibble remarks, ``string theory cosmologists have discovered cosmic strings lurking everywhere in the undergrowth''.

Therefore cosmic strings open a door to string theory; if cosmic strings could be observed it would provide the first evidence for string theory, as it is very important. At present great efforts have been put in searching observational evidence for cosmic strings, including the detections of gravitational wave radiations of string cusps and loops by the LIGO and LISA projects, Comic Microwave Background(CMB) B modes  by the Planck Surveyor Mission and BICEP2, etc \cite{Cohen:2010xd,Henrot-Versille:2014jua,Moss:2014cra,Ade:2013sjv,Ade:2013xla,Aasi:2013vna}.

In this paper we propose a generating mechanism for cosmic string structures, in terms of quintessence field of the dark energy. Dark energy has been thought to be a candidate to account for accelerating expansion and flatness of the universe; current measurements indicate that dark energy contributes 68.3\% of the total energy of the present observable universe \cite{Ade:2013sjv}. There are various models for dark energy such as quintessence, phantom and quintom \cite{Witterich:1988,Steinhardt:1999nw,Feng:2004ad}. In this paper we consider a $U(1)$ complex scalar model of quintessence acting as a background field of the universe,
\begin{equation}
\psi (x)=\phi^1(x)+i\phi^2(x), \hspace*{10mm} \phi^1,\phi^2 \in \mathbf{R},
\end{equation}
where $x=\left(\mathbf{x},x^0\right)$ denotes the coordinates of the Riemann-Cartan manifold $\mathbf{U}^4$ of the early universe.

In the following Sect.2 it will be shown that the singular distributions of this complex scalar field $\psi(x)$ are able to give rise to cosmic string structures. In Sect.3, after demonstrating the weakness of the traditional method of (self-)linking numbers, we will make use of the Chern-Simons type topological charge to construct the Kauffman bracket knot polynomial which is topologically a much more powerful invariant than the linking numbers. Then in Sect.4, examples of Hopf links, trefoil knots, figure-8 knots, Whitehead links and Borromean rings will be presented for reader convenience. Finally Sect.5 of Conclusion and Discussions will complete this paper.


\section{Cosmic strings constructed from scalar field}

The quintessence model of dark energy is given by the following minimal coupling between gravity and the background field \cite{Witterich:1988}

\begin{equation}
S=\int d^4x\sqrt{-g}\left[-\frac{1}{2}g^{\mu\nu}\partial_\mu \psi\partial_\nu\psi-V(\psi)\right], \hspace*{10mm} \mu,\nu = 1,2,3,0, \label{Actn}
\end{equation}
where $V(\psi)$ is a dark energy potential and model-dependent. 
The equation of motion reads
\begin{equation}
\partial_{\mu}\partial^{\mu}\psi-V'(\psi)=0. \label{EoM}
\end{equation}

With a given a boundary condition that delivers the topological information of the base manifold, one is about to find a solution $\psi(x)$ to eq.\eqref{EoM}. For this field $\psi$ distributed on the manifold, we are able to define a geodesic in the sense of a parallel field condition:
\begin{equation}
D_\mu \psi= \partial_\mu \psi -i\alpha A_\mu \psi=0, \label{Dmupsi}
\end{equation}
where $\alpha=\sqrt{\hbar G/c^3}$ is a coupling constant and $D_\mu$ a covariant derivative. $A_\mu$ is a $U(1)$ gauge potential (i.e., a connection of principal bundle) induced by $\psi$ through eq.\eqref{Dmupsi}, as long as $\psi$ is regarded as a section of an associate bundle on the manifold. $A_\mu$ is solved out from \eqref{Dmupsi} as
\begin{equation}\label{Amu}
A_{\mu}=\frac{\alpha}{2\pi}\frac{1}{2i\psi^*\psi}
\left(\psi^*\partial_{\mu}\psi-\partial_{\mu}\psi^*\psi\right) .
\end{equation}
Eq.\eqref{Amu} takes the same form of the velocity field in quantum mechanics, thanks to the London assumption of superconductivity \cite{duan:2002}.
The gauge field strength defined from $A_{\mu}$ is
\begin{equation}
F_{\mu\nu}=\partial_{\mu}A_{\nu}-\partial_{\nu}A_{\mu}.
\end{equation}

In order to derive topological defects we introduce a two-dimensional unit vector $n^a$ from $\phi^{1,2}$:
\begin{equation}
n^a=\frac{\phi^a}{\left\| \phi \right\|},\hspace*{15mm}a=1,2;~\left\| \phi \right\|^2=\phi^a\phi^a=\psi^*\psi.
\end{equation}
Thus
\begin{equation}
A_\mu=\frac{\alpha}{2\pi}\epsilon_{ab}n^a\partial_\mu n^b,
\hspace*{20mm} F_{\mu\nu}=\frac{\alpha}{2\pi}2\epsilon_{ab}\partial_\mu n^a\partial_\nu n^b.
\end{equation}
Introducing a topological tensor current \cite{Duan:1999tw},
\begin{equation}\label{Lgr}
j^{\mu\nu}=\frac{1}{2}\frac{1}{\sqrt{-g}}\epsilon^{\mu\nu\lambda\rho}F_{\lambda\rho},
\end{equation}
$F_{\mu\nu}$ can be re-expressed as a $\delta$-function form,
\begin{equation} \label{jmudelta}
j^{\mu\nu}=\frac{\alpha}{\sqrt{-g}}\delta^2(\mathbf{\phi})
D^{\mu\nu}\left(\frac{\phi}{x}\right),
\end{equation}
where Jacobian $\epsilon^{ab}D^{\mu\nu}\left(\frac{\phi}{x}\right) =
\epsilon^{\mu\nu\lambda\rho}\partial_{\lambda}\phi^a\partial_{\rho}\phi^b$. In deducing \eqref{jmudelta} the following relations also apply:
\[
\partial_{\mu}n^a=\frac{\partial_{\mu}\phi^a}{\left\|\phi \right\|}+\phi^a \partial_{\mu}\frac{1}{\left\|\phi \right\|},\hspace*{15mm}\partial_a\partial_a\ln \left\| \phi \right\|=2\pi\delta^2(\mathbf{\phi}).
\]

In the light of the topological tensor current $j^{\mu\nu}$ a Nambu-Goto action is constructed as \cite{Nielsen:1973}
\begin{equation} \label{NamGotoActn1}
S=\int_{\mathbf{U}^4}L \sqrt{-g}d^4x, \hspace*{10mm} \text{with} \hspace*{10mm} L=\frac{1}{\alpha}\sqrt{\frac{1}{2}g_{\mu\lambda}g_{\nu\rho}j^{\mu\nu}j^{\lambda\rho}}.
\end{equation}
Substituting \eqref{jmudelta} into \eqref{NamGotoActn1} we arrive at an important result,
\begin{equation}
L=\frac{1}{\sqrt{-g}}\delta^2(\mathbf{\phi}).
\end{equation}
Given that
\begin{equation}
\delta^2(\mathbf{\phi}) \left\{
             \begin{array}{lll}
              =0, & & \text{iff }\phi^a\neq 0; \\
              \neq 0, & & \text{iff }\phi^a= 0,
             \end{array}
        \right. \hspace*{15mm} a=1,2,
\end{equation}
the zero points of the $\phi^a$ field,
\begin{equation} \label{phi12eqn}
\phi^1(x^1,x^2,x^3,x^0)=0,\hspace*{15mm} \phi^2(x^1,x^2,x^3,x^0)=0,
\end{equation}
outline the nonzero evaluations of $L$ and indicate the presence of topological defects. According to the implicit function theorem, under the regular condition $D^{\mu\nu}\left(\frac{\phi}{x}\right)  \neq 0$, the two coupled equations of eq.\eqref{phi12eqn} have the following general solutions:
\begin{equation} \label{stringsolns}
P_k:\hspace*{3mm} x^\mu=x^\mu_k(u^I),\hspace*{15mm} k=1,2,\cdots,N;~I=1,2.
\end{equation}
which are $N$ isolated 2-dimensional singular submanifolds in the four dimensional spacetime, with $u^1,u^2$ being the intrinsic parameters. Thus, we achieve a model for cosmic strings by regarding $P_k$'s as the world-sheets swept by $N$ line defects, $k=1,2,\cdots,N$, with $u^1=s$ as the spatial parameter and $u^2=t$ the temporal one, respectively.

The Lagrangian $L$ then reads
\begin{equation} \label{Lagstrings}
L=\frac{1}{\sqrt{-g}}\sum_{k=1}^N W_k \int_{P_k}
\delta^4\left(x^{\mu} - x^{\mu}_k(s,t)\right) \sqrt{g_k} d^2u,
\end{equation}
where $g_k = \det \left[g_{IJ}\right]$, with $g_{IJ}$ as the intrinsic metric of the submanifold $P_k \left(u^I\right)$. $W_k$ is the topological charge of $P_k$, carrying the meaning of winding numbers, $W_k=\beta_k \eta_k$, with $\beta_k$ being the Hopf index and $\eta_k = \pm 1$ the Brouwer degree. Thus, the action can be expressed as a sum over the actions of the individual world-sheets,
\begin{equation} \label{actnstrings}
S=\sum_{k=1}^N W_k S_k \hspace*{10mm} \text{with}
\hspace*{10mm} S_k=\int_{P_k}\sqrt{g_k} d^2u,
\end{equation}
$S_k$ carrying the meaning of the area of $P_k$.

The spatial component of the current $j^{\mu\nu}$ is given by
\begin{equation} \label{topcurrspatial}
j^i=j^{0i}=\frac{\alpha}{\sqrt{-g}}\sum_{k=1}^{N}W_k\int_{l_k}
\frac{dx^i}{ds}\delta^3 \left(\mathbf{x}-\mathbf{x}_k (s)\right) ds, \hspace*{10mm} i= 1,2,3.
\end{equation}
$\frac{dx^i}{ds}$ gives the velocity of the $k$th cosmic string,
\begin{equation} \label{velocitystring}
\frac{dx^i}{ds} = \frac{D^i \left(\frac{\mathbf{\phi}}{x}\right)}{D \left(\frac{\mathbf{\phi}}{u}\right)}.
\end{equation}
The motion of a string is governed by the evolution equation:
\begin{equation} \label{stringevolutioneqn}
\frac{1}{\sqrt{g_k}}\frac{\partial}{\partial u^I}\left(\sqrt{g_k} g^{IJ} \frac{\partial x^{\mu}}{\partial u^J} \right)
+ g^{IJ} \Gamma^{\mu}_{\nu \lambda} \frac{\partial x^{\nu}}{\partial u^I} \frac{\partial x^{\lambda}}{\partial u^J} =0,
\hspace*{10mm} \mu,\nu=1,2,3,0;~I,J=1,2.
\end{equation}


\section{Topological invariants for knotted cosmic strings}

An ensemble of topological defects form a complex system, whose complexity is measured by its tangledness. Usually topological complexity of the system is strongly relevant to its energy and other dynamical properties: higher complexity corresponding to higher free energy or entropy, and energy release taking place during intercommutation collisions and complexity-reducing cascades \cite{Kleckner:2015}. Classification and characterization of topological complexity are at the center of the study of cosmic strings.

Hereinafter, we will investigate topology of cosmic strings from the mathematical point of view by studying closed strings; open strings ending at infinity will be treated as closed loops via compacticification. Thus, a string is an embedding map from a circle to the $3$-dimensional space, $\gamma : S^1 \rightarrow \mathbf{R}^3$.


\subsection{(Self-)linking numbers for knots}


A knotted string with finite length has finite energy from the viewpoint of soliton study. In 1997 Faddeev and Niemi proposed a soliton model beyond the classical
action of low energy Yang-Mills gauge theory, its solution providing the first $3$-dimensional topologically stable knotted soliton solution with finite energy \cite{Niemi:1997}. This model has found wide applications in physics, including trefoil knots in real fluids, protein folding in molecular biology,
topological insulators exhibiting knotted 3D electronic band structure in condensed matter, etc. \cite{Kleckner:2013,Shakhnovich:2011,Moore:2010}.

In the Faddeev-Niemi model the tangledness of knotted solitons is characterized by the topological charge
\begin{equation} \label{eq:QCSdefn}
Q=\left(\frac{2\pi}{\alpha}\right)^2\frac{1}{4\pi}\int_\Omega \epsilon^{ijk}A_iF_{jk}d^3x = \frac{2\pi}{\alpha^2}\int_\Omega A_i j^{i} \sqrt{-g} d^3x, \hspace*{15mm} \epsilon^{ijk}=\epsilon^{0ijk},
\end{equation}
where $\Omega$ is the special volume. $A_i$ and $F_{ij}$ are the spatial components of $A_{\mu}$ and $F_{\mu \nu}$, respectively. Mathematically, $Q$ is a Hopf map, $\pi_3(S^2)=\mathbf{Z}$. In the context of fluid mechanics it corresponds to an important topological invariant, helicity: $H = \int_\Omega \mathbf{u} \cdot \mathbf{\omega} d^3x $. Here $\mathbf{u}$ is the velocity of the fluid and $\mathbf{\omega}$ the vorticity, $\mathbf{\omega} = \nabla \times \mathbf{u}$ \cite{Moffatt:1969}.
Bekenstein pointed out that this helicity can be transplanted into the study of cosmic strings \cite{Bekenstein:1992}.

Moffatt and Ricca successfully derived an algebraic expression for $H$ as a sum of all (self-)linking numbers of knotted fluid knots \cite{Ricca:1992}; the strict verification of this formula in the context of cosmic strings was given by Duan and Liu \cite{Liuprd:2003,LiuJHEP2004}:
\begin{equation}\label{CScharge2}
Q=2\pi\left[\sum_{k=1}^N W^2_k SL(\gamma_k)+2\sum_{k,l=1(k< l)}^N W_kW_l Lk(\gamma_k,\gamma_l) \right],
\end{equation}
where $SL(\gamma_k)$ is the self-linking number of the $k$th knotted string, and $LK(\gamma_k,\gamma_l)$ the Gauss (mutual) linking number between the $k$th and $l$th strings. The (self-)linking numbers are easy to obtained by counting the degree of each crossing site. By eq.\eqref{CScharge2}, the initial triple integral, eq.\eqref{QCSdefn}, is advantageously turned into simple algebraic counting.

However, from the knot theoretical point of view, (self-)linking numbers are weak knot invariants which fail to distinguish typical topology of knots/links, as shown in Figs. \ref{Example1} and \ref{Example2}:
\begin{figure}[H]
\centering
\subfigure[]{\includegraphics[width=3cm]{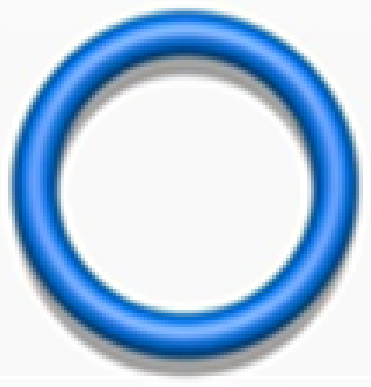}}
\hspace*{20mm}
\subfigure[]{\includegraphics[width=3cm]{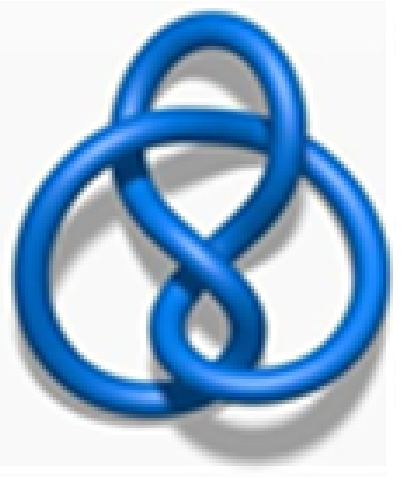}}
\hspace{5ex}
\caption{Different topological configurations which have the same self-linking number $0$: (a) a topologically trivial circle; (b) a topologically non-trivial figure-8 knot.}
\label{Example1}
\end{figure}

\begin{figure}[H]\centering
\subfigure[]{\includegraphics[width=3cm]{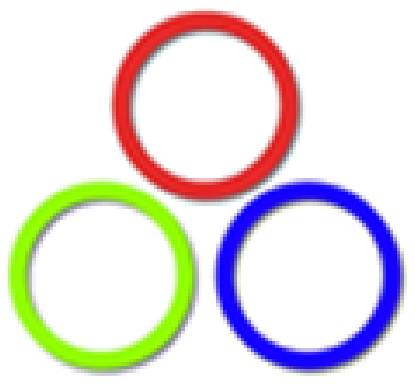}}
\hspace*{10mm}
\subfigure[]{\includegraphics[width=3cm]{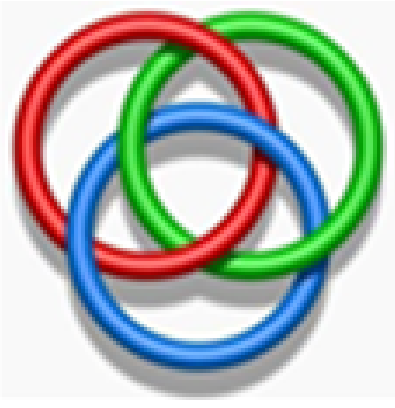}}
\hspace*{10mm}
\subfigure[]{\includegraphics[width=3cm]{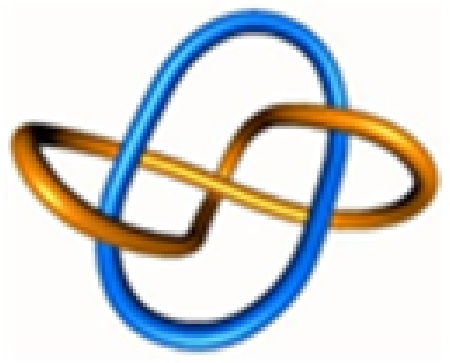}}
\caption{Different topological configurations which have the same linking number $0$: (a) three disjoint trivial circles; (b) the Borromean rings; (c) the Whitehead links.}
\label{Example2}
\end{figure}

Therefore, in order to distinguish and characterize topology of cosmic strings we need to develop new and stronger tools of topological invariants. In the following text our starting point will still be the topological charge $Q$ of eq.\ref{eq:QCSdefn}, due to a remarkable fact that $Q$ is an abelian Chern-Simons action. Chern-Simons theory is well known to be the most important topological quantum field theory in three dimensions, which provides a field theoretical framework for knot theory \cite{Witten:1989}. It is strongly relevant to knot topological invariants such as the Jones, HOMFLYPT and other knot polynomials, the Vassiliev finite type invariants via the Kontsevich integral, Khovanov homology, and so on \cite{Birman:1993,Lin:2000,Barnatan:1995}.

In the next two subsections we will start from the Chern-Simons type topological charge $Q$ to derive the Kauffman bracket polynomial by constructing the latter's skein relations.


\subsection{Exponential form of topological charge $Q$}

Substituting the $\delta$-function expression of the current $j^i$, eq.\eqref{topcurrspatial}, into eq.\eqref{eq:QCSdefn} we obtain
\[
Q=\frac{2\pi}{\alpha}\sum_{k=1}^N W_k\oint_{\gamma_k}A_i d x^i.
\]
For convenience, in this paper we set $W_k=1$ and rescale $\alpha =1$ (\textit{for the cases $W_k , \alpha \neq 1$ one can formally keep $W_k$ and $\alpha$ intact}). Thus,
\begin{equation}\label{Qcircintgl}
Q=2\pi\sum_{k=1}^N \oint_{\gamma_k}A_i d x^i = 2\pi \oint_{L}A_i d x^i,
\hspace*{15mm}
L = \bigoplus\limits_{k=1}^{N} \gamma_k,
\end{equation}
where $L$ is the link composed of all $\gamma_k$'s.
Eq.\eqref{Qcircintgl} means that when thin cosmic strings are considered, $Q$ can be expressed as a circular integral in the configuration space.

In this paper our main proposal is to study the exponential form of $Q$:
\begin{equation} \label{ExpForm-1}
e^{\frac{1}{2\pi}Q(L)} = e^{\oint_{L}A_i d x^i}.
\end{equation}
We argue that this form is able to provide knot polynomial topological invariants such as the Kauffman bracket polynomial (see below). The advantages of this form are the following:
\begin{itemize}
  \item \textsf{Additivity of the line integral}: The link $L$ is divided into the sum of a few strands $S_1,S_2,S_3,\cdots$ in the configuration space, $L = S_1 \oplus S_2 \oplus S_3 \oplus \cdots$, as shown in Fig.\ref{Strandecompn}.
  \begin{figure}[H]
      \centering
      \includegraphics[width=0.25\textwidth]{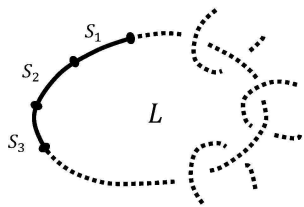}
      \caption{The link $L$ is additive, i.e., it can be treated as the sum of all the strands $S_1,S_2,S_3,\cdots$.}
      \label{Strandecompn}
  \end{figure}
  Then the line integral over the path $L$ has the feature of additivity:
  \begin{equation}\label{Strandecompn-1}
    \oint_L = \int_{S_1} + \int_{S_2} + \int_{S_3} + \cdots .
  \end{equation}

  \item \textsf{Factorization of the exponential form}: The addition in the power of the exponential leads to
  \begin{equation}\label{factornExp}
    e^{\oint_L} = e^{\int_{S_1}}e^{\int_{S_2}}e^{\int_{S_3}}\cdots.
  \end{equation}
  This provides us a pathway to construct the formal parameter $a$ in the skein relations of the Kauffman bracket polynomial.
\end{itemize}

\noindent We introduce a bracket symbol to denote the exponential of the line integral over a link $L$:
\begin{equation}\label{DefBracketSymbol}
\left\langle L \right\rangle= e^{\oint_{L}A_i d x^i}.
\end{equation}
An immediate application is the case that $L$ is a trivial circle $\PICcircle$:
\begin{equation}\label{1stskeinRel}
  \left\langle \PICcircle \right\rangle= e^{\oint_{\PICcircle}A_i d x^i}
  = e^{\frac12\iint_{\Sigma}F_{ij} dx^i \wedge dx^j} = e^0 = 1,
\end{equation}
where the Stokes' theorem applies.


\subsection{Kauffman bracket knot polynomial topological invariant} \label{SectKauffcompn}

\begin{figure}[H]
\centering
\includegraphics[width=0.82\textwidth]{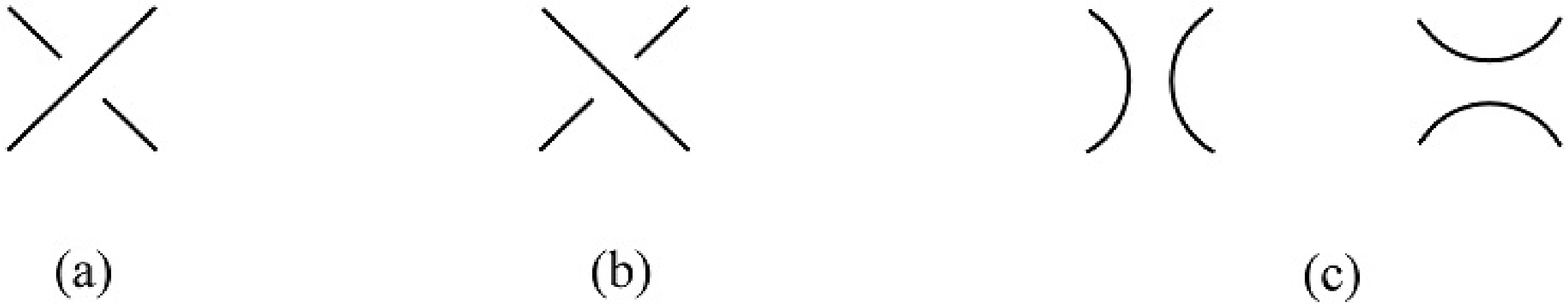}
\caption{The diagrams of four almost-the-same un-oriented knots/links, which differ only at one particular crossing site. (a) Over-crossing, denoted as $L_+$. (b) Under-crossing, denoted as $L_-$. (c) Non-crossing: (\textit{left}) $L_0$, containing two branches, left and right;  (\textit{right}) $L_{\infty}$, containing two branches, up and down.}
\label{crossingstates}
\end{figure}

To keep in accordance with the knot theoretical routine, let us start from examining the three basic states at a crossing site: the over-, under- and non-crossings, as shown in Fig\ref{crossingstates}.

\begin{figure}[H]
\centering
\includegraphics[width=0.78\textwidth]{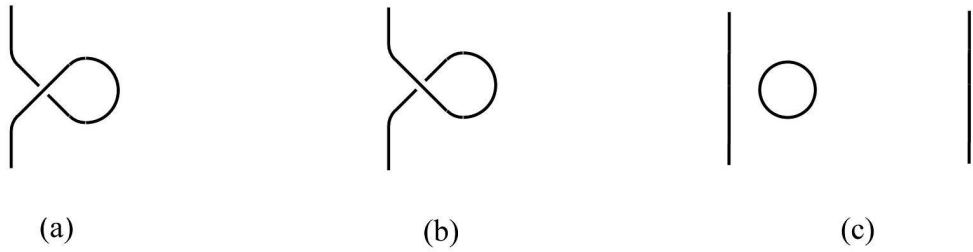}
\caption{Configurations obtained by joining the two right ends of each one of Fig.\ref{crossingstates}. (a) A writhe denoted as $\alpha_+$, obtained by joining the two right ends of $L_+$. (b) A writhe denoted as $\alpha_-$, obtained by joining the ends of $L_-$. (c) \textit{Left}: A disjoint union denoted as $\alpha_{0}$, obtained by joining the two right ends of $L_0$;  \textit{Right}: The rest of the link, denoted as $\alpha_{\infty}$, obtained by joining the two right ends of $L_{\infty}$.}
\label{writhestetes}
\end{figure}

\noindent An example is that, if joining the two right ends in each configuration of Fig.\ref{crossingstates}, we obtain the ones of Fig.\ref{writhestetes}: (a) $\alpha_+$, (b) $\alpha_-$, and (c) $\alpha_0 = \PICline \cup \PICcircle$ and $\alpha_{\infty} = \PICline$.

\noindent Furthermore, for convenience, we introduce the following notations for two important writhing loops:
\[
\gamma_+ = \PICclosegammaplus,\hspace*{15mm}\gamma_- = \PICclosegammaminus.
\]

The over- and under-crossings $L_{+}$ and $L_{-}$, as integration paths in the sense of eq.\eqref{Qcircintgl}, can be locally decomposed in different ways in the configuration space. Taking $L_+$ for example, it has two decomposing channels, the left-right (LR) and the up-down (UD), as showed in Fig. \ref{crossing} (for details see \cite{Liu:2012,Liu:2014}):
\begin{figure}[H]
\centering
\includegraphics[width=0.55\textwidth]{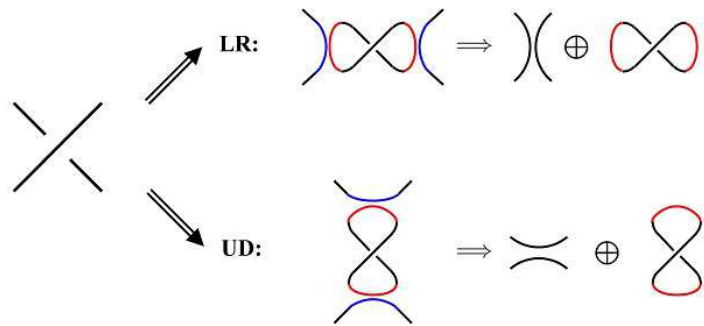}
\caption{Two decomposing channels for the over-crossing $L_+$: left-right (LR) and up-down (UD) splittings. [\textit{Upper row}, LR]: By adding local imaginary red \& blue strands which can cancel each other, $L_+$ is split into a non-crossing $L_0$ plus a writhing loop $\gamma_+$. [\textit{Lower row}, UD]: By adding local imaginary strands, $L_+$ is split into a non-crossing $L_{\infty}$ plus a writhing loop $\gamma_-$.}
\label{crossing}
\end{figure}

\noindent Adopting ergodic statistical hypothesis, we argue that these two channels should have equal contributions to $e^{\frac{1}{2\pi}Q\left(L_+\right)}$ when modifying the integration paths of $L_+$:
\begin{eqnarray}
&& e^{\frac{1}{2\pi}Q(L_+)} = \left\langle L_+ \right\rangle = \left\langle L_+ \text{-- LR}\right\rangle
+ \left\langle L_+ \text{-- UD}\right\rangle , \label{D0}
\\
\text{where} \hspace*{10mm}
&& \left\langle L_+ \text{-- LR}\right\rangle = e^{\oint_{L_0 \oplus \gamma_+}}
= e^{\oint_{L_0}}e^{\oint_{\gamma_+}}, \label{D0-1}
\\
&& \left\langle L_+ \text{-- UD}\right\rangle = e^{\oint_{L_{\infty} \oplus \gamma_-}}
= e^{\oint_{L_{\infty}}}e^{\oint_{\gamma_-}}. \label{D0-2}
\end{eqnarray}
Denoting $e^{\oint_{\gamma_{+,-}}}$ as
\begin{eqnarray}
e^{\oint_{\gamma_+}} &=& \exp \left(\oint_{\PICclosegammaplus} A_i d x^i \right) = a, \label{D0-3}\\
e^{\oint_{\gamma_-}} &=& \exp \left(\oint_{\PICclosegammaminus} A_i d x^i \right) =
\exp \left(- \oint_{\PICclosegammaplus} A_i d x^i \right) = a^{-1}, \label{D0-4}
\end{eqnarray}
eq.\eqref{D0} becomes
\begin{equation}\label{D1}
 \left\langle \PICunorientpluscross \right\rangle =
 a \left\langle \PICunorientLRsplit \right\rangle +
 a^{-1} \left\langle \PICunorientUDsplit \right\rangle,
\end{equation}
where
\begin{equation}
 \left\langle \PICunorientpluscross \right\rangle = e^{\oint_{L_+}},
  \hspace*{15mm}
 \left\langle \PICunorientLRsplit \right\rangle = e^{\oint_{L_0}},
 \hspace*{15mm}
 \left\langle \PICunorientUDsplit \right\rangle = e^{\oint_{L_{\infty}}}.
\end{equation}
Similarly, $e^{\frac{1}{2\pi}Q(L_-)}$ has the decomposition
\begin{equation}\label{D2}
 \left\langle \PICunorientminuscross \right\rangle =
 a^{-1} \left\langle \PICunorientLRsplit \right\rangle +
 a \left\langle \PICunorientUDsplit \right\rangle, \hspace*{10mm}
 \text{with~~~} \left\langle \PICunorientminuscross \right\rangle = e^{\oint_{L_-}}.
\end{equation}

The evaluation of $a$ gives the directional writhing number of \PICclosegammaplus: $a = e^{\lambda}$, where $\lambda$ ranges within $[0,1]$, reflecting the direction to observe the writhe. The statistical average of $\lambda$ reads $\bar{\lambda} = \frac12$, thus $\bar{a} = e^{\frac12}$.

Eqs.\eqref{D1} and \eqref{D2} account for the two skein relations describing crossings. To obtain another skein relation describing the disjoint union $\left\langle \PICline \cup \PICcircle \right\rangle$, we notice the following splitting decompositions:
\begin{equation}
e^{\oint_{\alpha_+}} = \exp\left(\oint_{\PICline \cup \PICcircle} \right)
\exp\left(\oint_{\gamma_+} \right), \hspace*{10mm}
e^{\oint_{\alpha_-}} = \exp\left(\oint_{\PICline \cup \PICcircle} \right)
\exp\left(\oint_{\gamma_-} \right), \label{aEvaluation}
\end{equation}
illustrated by
\begin{figure}[H]
\centering
\includegraphics[width=10cm]{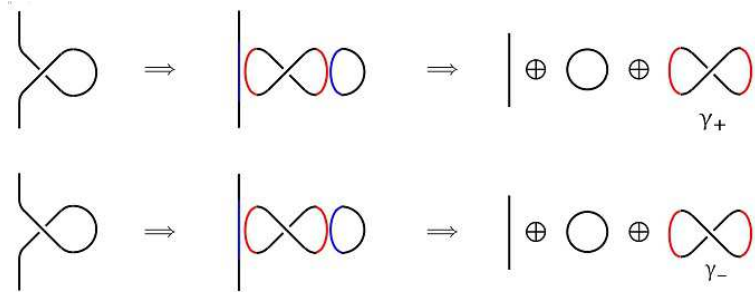}
\caption{(a) \textit{Upper row}: splitting of the writhe $\alpha_+$. (b) \textit{Lower row}: splitting of the writhe $\alpha_-$.}
\label{Deomposition}
\end{figure}

\noindent This means we have two channels to derive $\left\langle\PICline \cup \PICcircle \right\rangle$ as shown in Fig.\ref{linecircle}:
\begin{figure}[H]
\centering
\includegraphics[width=0.45\textwidth]{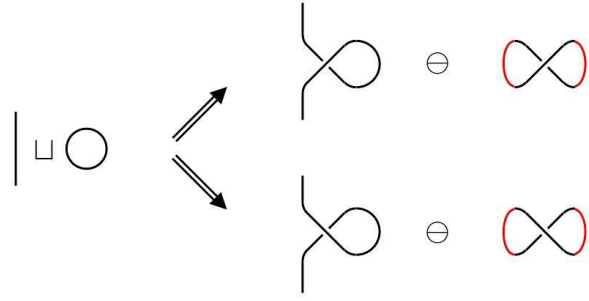}
\caption{Two ways to achieve the disjoint union in the configuration space: (\textit{upper row}) by imaginarily subtracting $\gamma_+$ from $\alpha_+$; and (\textit{lower row}) by imaginarily subtracting $\gamma_-$ from $\alpha_-$.}
\label{linecircle}
\end{figure}

\noindent As per the ergodic statistical hypothesis these two channels should have the equal contribution to $\left\langle\PICline \cup \PICcircle \right\rangle$; hence,
 \begin{equation} \label{3rdskeinRel-1}
\left\langle \PICline \cup \PICcircle \right\rangle
= a^{-1} \left\langle \PICopengammaplus \right\rangle
+ \left(a^{-1}\right)^{-1} \left\langle \PICopengammaminus \right\rangle
= a^{-1} \left\langle \PICopengammaplus \right\rangle
+ a \left\langle \PICopengammaminus \right\rangle.
\end{equation}

On the other hand, $\PICopengammaplus$ (resp., $\PICopengammaminus$) can be obtained by joining the two right ends of $\PICunorientpluscross$ (resp., \PICunorientminuscross). Thus eqs.\eqref{D1} and \eqref{D2} yields
\begin{eqnarray}
  \left\langle \PICopengammaplus \right\rangle &=&
   a \left\langle \PICline \cup \PICcircle \right\rangle +
 a^{-1} \left\langle \PICline \right\rangle, \label{3rdskeinRel-2}\\
    \left\langle \PICopengammaminus \right\rangle &=&
   a^{-1} \left\langle \PICline \cup \PICcircle \right\rangle +
 a \left\langle \PICline \right\rangle, \label{3rdskeinRel-3}
\end{eqnarray}
Combining eqs.\eqref{3rdskeinRel-1}, \eqref{3rdskeinRel-2} and \eqref{3rdskeinRel-3}, we arrive at
\begin{equation}\label{3rdskeinRel}
  \left\langle \PICline \cup \PICcircle \right\rangle
  = \left( -a^{-2} -a^2\right) \left\langle \PICline \right\rangle .
\end{equation}

Then, in summary, we have achieved the skein relations of the Kauffman bracket polynomial, i.e., eqs.\eqref{1stskeinRel}, \eqref{D1}, \eqref{D2} and \eqref{3rdskeinRel}:
\begin{eqnarray}
  && \left\langle \PICcircle \right\rangle = 1, \label{KaufBraSkeinRel-1} \\
  && \left\langle \PICunorientpluscross \right\rangle =
    a \left\langle \PICunorientLRsplit \right\rangle +
    a^{-1} \left\langle \PICunorientUDsplit \right\rangle,
    \hspace*{10mm}
     \left\langle \PICunorientminuscross \right\rangle =
    a^{-1} \left\langle \PICunorientLRsplit \right\rangle +
    a \left\langle \PICunorientUDsplit \right\rangle,
    \label{KaufBraSkeinRel-2}\\
   && \left\langle \PICline \cup \PICcircle \right\rangle
    = \left( -a^{-2} -a^2\right) \left\langle \PICline \right\rangle . \label{KaufBraSkeinRel-3}
\end{eqnarray}

An additional remark is that the Jones polynomial for oriented knots/links is able to be constructed from the Kauffman bracket polynomial for un-oriented ones. See Refs.\cite{KauffmanBook1987,Liu:2012}.

\section{Examples}

In this section the Kauffman bracket polynomials of some typical elementary configurations are presented for reader convenience.


\subsection{Disjoint union of trivial circles}

Let us start from considering the union of two circles. In terms of \eqref{KaufBraSkeinRel-3} we have
\begin{equation*}
  \left\langle \PICcircle \cup \PICcircle \right\rangle
    =  \left( -a^{-2} -a^2 \right) \left\langle \PICcircle \right\rangle
    =  -a^{-2} -a^2.
\end{equation*}
This result can be easily generalized to the case of $n$ circles:
\begin{equation} \label{DisUnionNcircles}
  \left\langle \overset{n\text{ copies}}{\overbrace{\PICcircle \cup \PICcircle \cup \cdots \cup \PICcircle}} \right\rangle
    = \left(- a^{-2} - a^2 \right)^{n-1}.
\end{equation}


\subsection{Hopf link, $H$}

\begin{figure}[H]
\centering
\includegraphics[width=8.5cm]{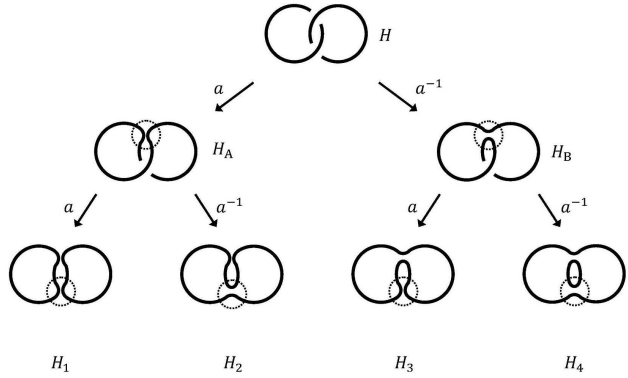}
\caption{Computation of the Kauffman bracket polynomial of a Hopf link.}
\label{PicHopf}
\end{figure}
Consider a Hopf link, denoted as $H$, as shown in the top row of Fig.\ref{PicHopf}. Applying the skein relation \eqref{KaufBraSkeinRel-2} to its upper crossing site, we have
\begin{equation}\label{HopfLink-1}
  \left\langle H \right\rangle = a \left\langle H_A \right\rangle
  + a^{-1} \left\langle H_B \right\rangle,
\end{equation}
where $H_A$ and $H_B$ are the two states in the middle row of Fig.\ref{PicHopf}. Further, continuing to apply \eqref{KaufBraSkeinRel-2} to the lower crossing site of $H_A$, we have
\begin{equation}\label{HopfLink-2}
  \left\langle H_A \right\rangle = a \left\langle H_1 \right\rangle
  + a^{-1} \left\langle H_2 \right\rangle,
\end{equation}
where $H_1$ and $H_2$ are two states in the bottom row of Fig.\ref{PicTrefoil}. Similarly, $H_B$ satisfies
\begin{equation}\label{HopfLink-3}
  \left\langle H_B \right\rangle = a \left\langle H_3 \right\rangle
  + a^{-1} \left\langle H_4 \right\rangle.
\end{equation}

Substituting \eqref{HopfLink-2} and \eqref{HopfLink-3} into \eqref{HopfLink-1} we have\begin{equation}\label{HopfLink-4}
  \left\langle H \right\rangle
  = aa \left\langle H_1 \right\rangle
  + aa^{-1} \left\langle H_2 \right\rangle
  + a^{-1} a \left\langle H_3 \right\rangle
  + a^{-1}a^{-1} \left\langle H_4 \right\rangle .
\end{equation}
This expression can be understood as follows. It contains four items respectively corresponding to the four states in the bottom row. Each item, say $aa \left\langle H_1 \right\rangle$, has two factors: $\left\langle H_1 \right\rangle$, the polynomial of the state; $aa$, the path from the top $H$ to the bottom $H_1$ via the middle $H_A$.

In the light of eq.\eqref{DisUnionNcircles}, one can see that $\left\langle H_1\right\rangle = -a^2 -a^{-2}$, since $H_1$ the disjointed union of two trivial circles. Similarly, the bracket polynomials of the other three states $H_2$, $H_3$ and $H_4$ are
\begin{center}
\begin{tabular}{|l|l|l|l|}
\hline
$H_1$ & $- a^2 - a^{-2}$ & $H_2$ & $1$   \\ \hline
$H_3$ & $1$                   & $H_4$ & $- a^2 - a^{-2}$   \\ \hline
\end{tabular}
\end{center}
Then, substituting these values into \eqref{HopfLink-4} we achieve the Kauffman bracket polynomial of the Hopf link:
\begin{equation}\label{HopfLinkBrackPoly}
  \left\langle H \right\rangle = - a^{-4} - a^4.
\end{equation}

The above method can be extended to a generic case. For a link $L$, its Kauffman bracket polynomial is given by \cite{KauffmanBook1987}
\begin{equation}\label{GenericKauffBraFormula}
  \left\langle L \right\rangle = \sum_{s} a^{\theta_0 (s)} a^{-\theta_{\infty} (s)}
  \left( -a^{-2} -a^2 \right)^{\left|s\right|-1},
\end{equation}
where the sum is done over all the states in the bottom row. For a state, say $s$, supposing it is the disjoint union of $\left| s \right|$ circles, and from the top $L$ to this state $s$ the path contains $\theta_0 (s)$ LR-splittings and $\theta_{\infty} (s)$ UD-splittings, the summand corresponding to $s$ reads: $a^{\theta_0 (s)} a^{-\theta_{\infty} (s)}
  \left( -a^{-2} -a^2 \right)^{\left|s\right|-1}$.


\subsection{Trefoil knots, $T^R$ and $T^L$}

\begin{figure}[H]
\centering
\includegraphics[width=0.65\textwidth]{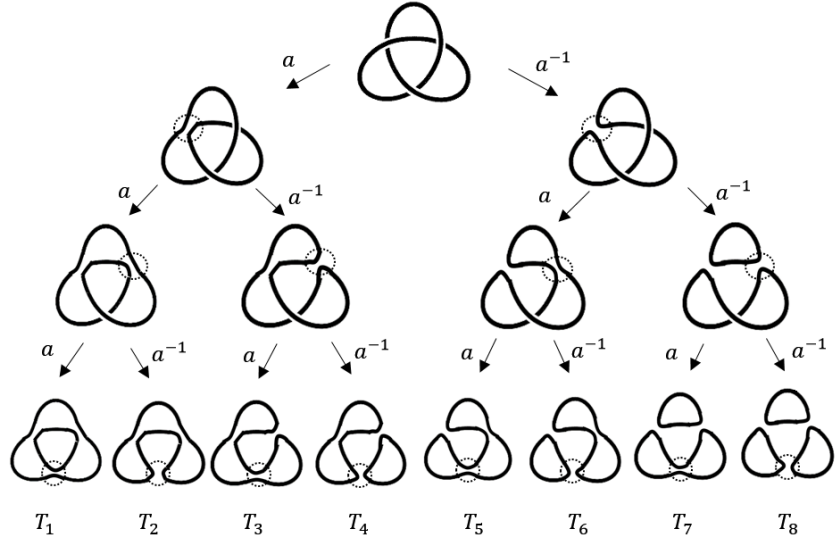}
\caption{Computation of the Kauffman bracket polynomial of a right-handed trefoil knot $T^R$.}
\label{PicTrefoil}
\end{figure}

In the light of eq.\eqref{GenericKauffBraFormula}, we have the bracket polynomials for each state in the bottom row of Fig.\ref{PicTrefoil}:
\begin{center}
\begin{tabular}{|l|l|l|l|l|l|}
\hline
$T_1$  & $-a^3(a^2+a^{-2})$ & $T_2$ & $a$ & $T_3$ & $a$ \\ \hline
$T_4$ & $-a^{-1}(a^2+a^{-2})$  & $T_5$ & $a$ & $T_6$ & $-a^{-1}(a^2+a^{-2})$ \\ \hline
$T_7$ & $-a^{-1}(a^2+a^{-2})$  & $T_8$ & $-a^{-3}(a^2+a^{-2})^2 $ & & \\ \hline
\end{tabular}
\end{center}
Summing up all the contributions of the states $T_1$---$T_8$, we achieve the Kauffman bracket for the right-handed trefoil knot $T^R$:
\begin{equation}
\left\langle T^R \right\rangle =a^{-7}-a^{-3}-a^5.
\end{equation}

Similarly, the polynomial of the left-handed trefoil knot $T^L$ (i.e., the mirror image of $T^R$) reads
\begin{equation}
\left\langle T^L \right\rangle =-a^{-5}-a^{3}+a^{7}.
\end{equation}
Obviously, the Kauffman bracket polynomial is able to distinguish the right- and left-handed trefoil knots.

\subsection{Figure-8 knot, $F^8$}

A Figure-8 knot, denoted as $F^8$, can be decomposed into the two states $A$ and $B$ in Fig.\ref{PicEight0}:
\begin{figure}[H]
\centering
\includegraphics[width=0.25\textwidth]{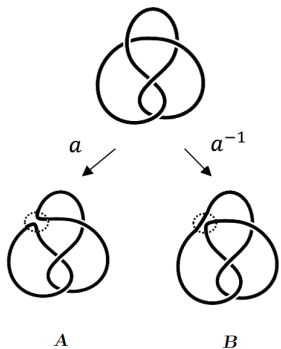}
\caption{Computation of the Kauffman bracket polynomial of the Figure-8 knot $F^8$: A knot $F^8$ (top row) can be decomposed into the states $A$ and $B$ (bottom row).}
\label{PicEight0}
\end{figure}

\noindent Thus
\begin{equation}\label{Fig8-1}
  \left\langle F^8 \right\rangle = a \left\langle A \right\rangle
  + a^{-1} \left\langle B \right\rangle.
\end{equation}
The further decomposition of $A$ is given in Fig.\ref{PicEight1}:
\begin{figure}[H]
\centering
\includegraphics[width=0.8\textwidth]{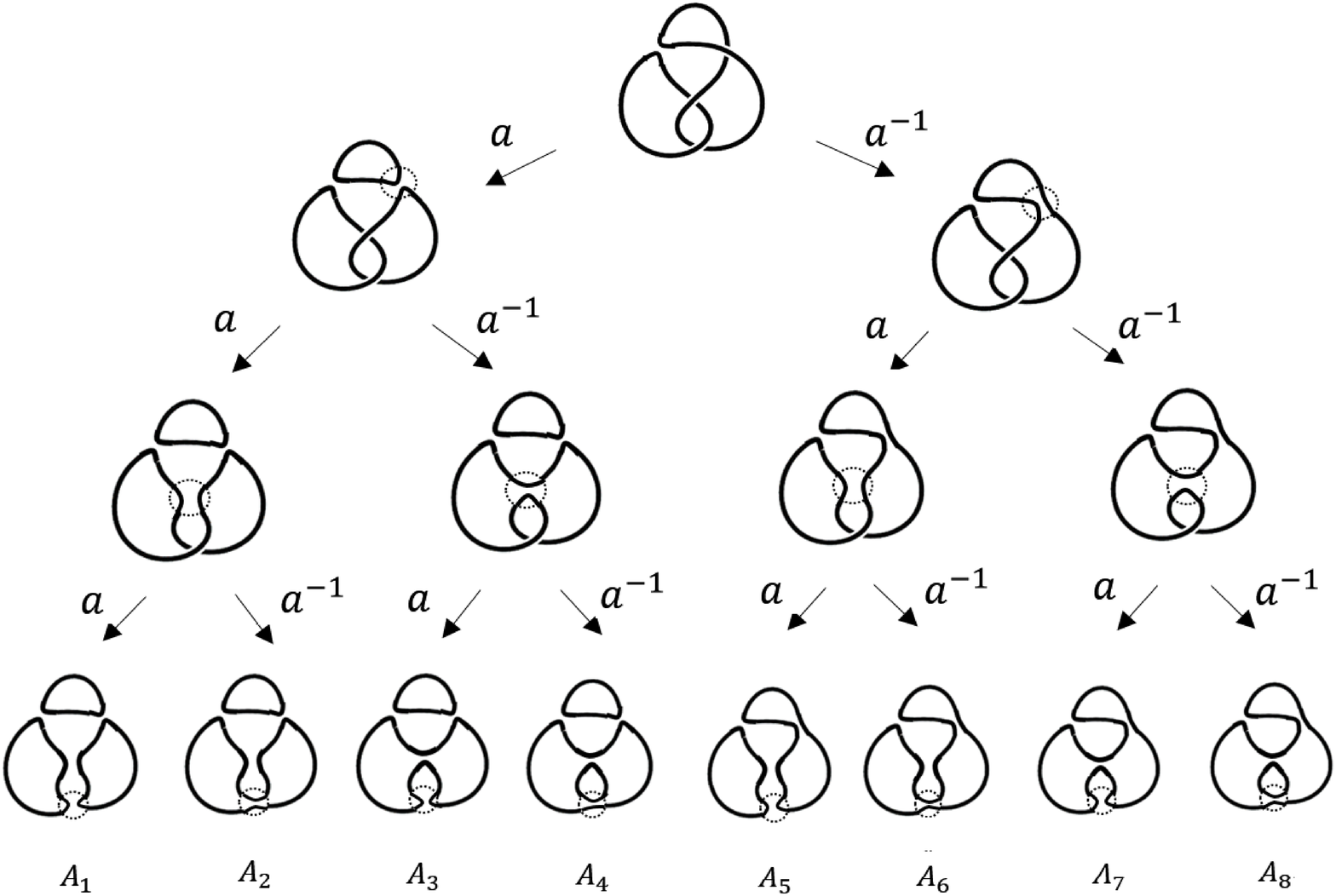}
\caption{Decomposition of the state $A$ of Fig\ref{PicEight0}.}
\label{PicEight1}
\end{figure}

\noindent where the polynomials for each of the states $A_1$---$A_8$ are

\begin{center}
\begin{tabular}{|l|l|l|l|l|l|l|l|}
\hline
$A_1$ & $a^4(a^{-2}+a^2)^2$ & $A_2$ &$-a^2(a^{-2}+a^2)$ & $A_3$ & $-a^2(a^{-2}+a^2)$ & $A_4$ & $(a^{-2}+a^2)^2$\\ \hline
$A_5$ & $-a^2(a^{-2}+a^2)$ & $A_6$ &1                    & $A_7$ & 1&$A_8$ & $-a^2(a^{-2}+a^2)$\\  \hline
\end{tabular}
\end{center}

The decomposition of $B$ is given in Fig.\ref{PicEight2}:
\begin{figure}[H]
\centering
\includegraphics[width=12cm]{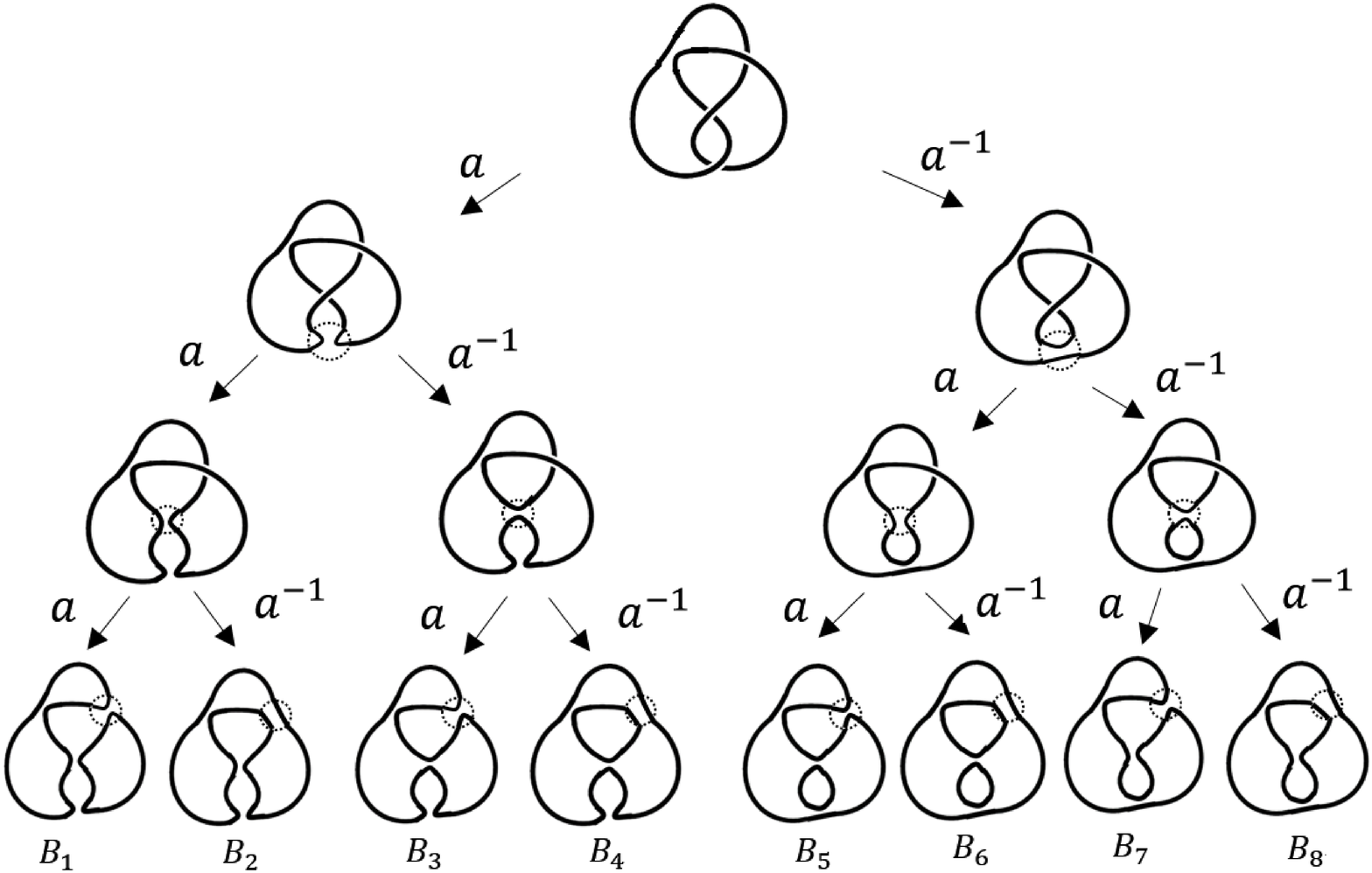}
\caption{Decomposition of the state $B$ of Fig\ref{PicEight0}.}
\label{PicEight2}
\end{figure}

\noindent where the polynomials for each of the states $B_1$---$B_8$ are

\begin{center}
\begin{tabular}{|l|l|l|l|l|l|l|l|}
\hline
$B_1$ & $-a^2(a^{-2}+a^2)$ & $B_2$ &1 & $B_3$ & 1&$B_4$ & $-a^2(a^{-2}+a^2)$\\ \hline
$B_5$ & $1 $ & $B_6$ &$-a^{-2}(a^{-2}+a^2)$  & $B_7$ & $-a^{-2}(a^{-2}+a^2)$ &$B_8$ & $-a^4(a^{-2}+a^2)^2$\\  \hline
\end{tabular}
\end{center}

Substituting states $A_1$---$A_8$ and $B_1$---$B_8$ into eq.\eqref{Fig8-1}, we achieve the Kauffman bracket polynomial for the Figure-8 knot:
\begin{equation}
\left\langle F^8 \right\rangle =a^{-8}-a^{-4}+1-a^4+a^8.
\end{equation}

An immediate conclusion is that the Kauffman polynomial is able to distinguish a trivial circle and a Figure-8 knot, the two topologically different configurations of Fig.\ref{Example1}.

\subsection{Borromean rings and Whitehead link}

The Kauffman polynomials of the Borromean rings and the Whitehead links are computed as
\begin{eqnarray}
&& \left\langle\text{Borromean rings}\right\rangle = -a^{-12} +3a^{-8} -2a^{-4} +4
 -2 a^4 +3 a^8 -a^{12}, \\
&& \left\langle\text{Whitehead links}\right\rangle = a^{-9} -a^{-5} +2a^{-1} -a^3 +2a^7 -a^{11}.
\end{eqnarray}
Meanwhile from eq.\eqref{DisUnionNcircles} we have the polynomial for three trivial circles
\begin{equation}
\left\langle\text{Three circles}\right\rangle = a^{-4}+2+a^4.
\end{equation}
Therefore the three topologically different configurations of Fig. \ref{Example2} can be successfully distinguished by the Kauffman bracket polynomial.

\section{Conclusion and Discussion}

In this paper cosmic strings in the early universe are achieved as one-dimensional topological defects in the $U(1)$ complex scalar quintessence field of dark energy.
Our starting point is the abelian Chern-Simons type topological charge $Q$. We propose to study its exponential form, and argue that the Kauffman bracket knot polynomial, a topological invariant much stronger than the traditional tool of (self-)linking numbers, can be constructed from this form. This opens a new door to investigation of structural complexity of extensive tangled cosmic strings in terms of classical field theory. Moreover, typical elementary examples of Hopf links, trefoil knots, figure-8 knot, Whitehead links and Borromean rings are presented for reader convenience.

\subsection*{Discussion: reconnection of cosmic strings}

In Subsection \ref{SectKauffcompn} we introduced a mathematical approach of breaking-reconnection, i.e., the technique of addition/subtraction of imaginary local paths together with ergodic statistical hypothesis. This provides a promising candidate for studying the physical reconnection processes of tangled cosmic strings within complexity-reducing cascades.

When moving towards each other, two cosmic strings probably cross and exchange strands, somehow like the production and annihilation of kink/anti-kink pairs.
This reconnection process has strong linkage to string compactification and coupling with certain relative velocities, intersection angles and string strengths (see \cite{Jackson:2004zg,Dubath:2007mf,Copeland:2009dk} for discussions of dynamical aspects of kinks in small and large scales).

\begin{figure}[H]
\centering
\subfigure[Intercommunication.]{
\includegraphics[width=0.32\textwidth]{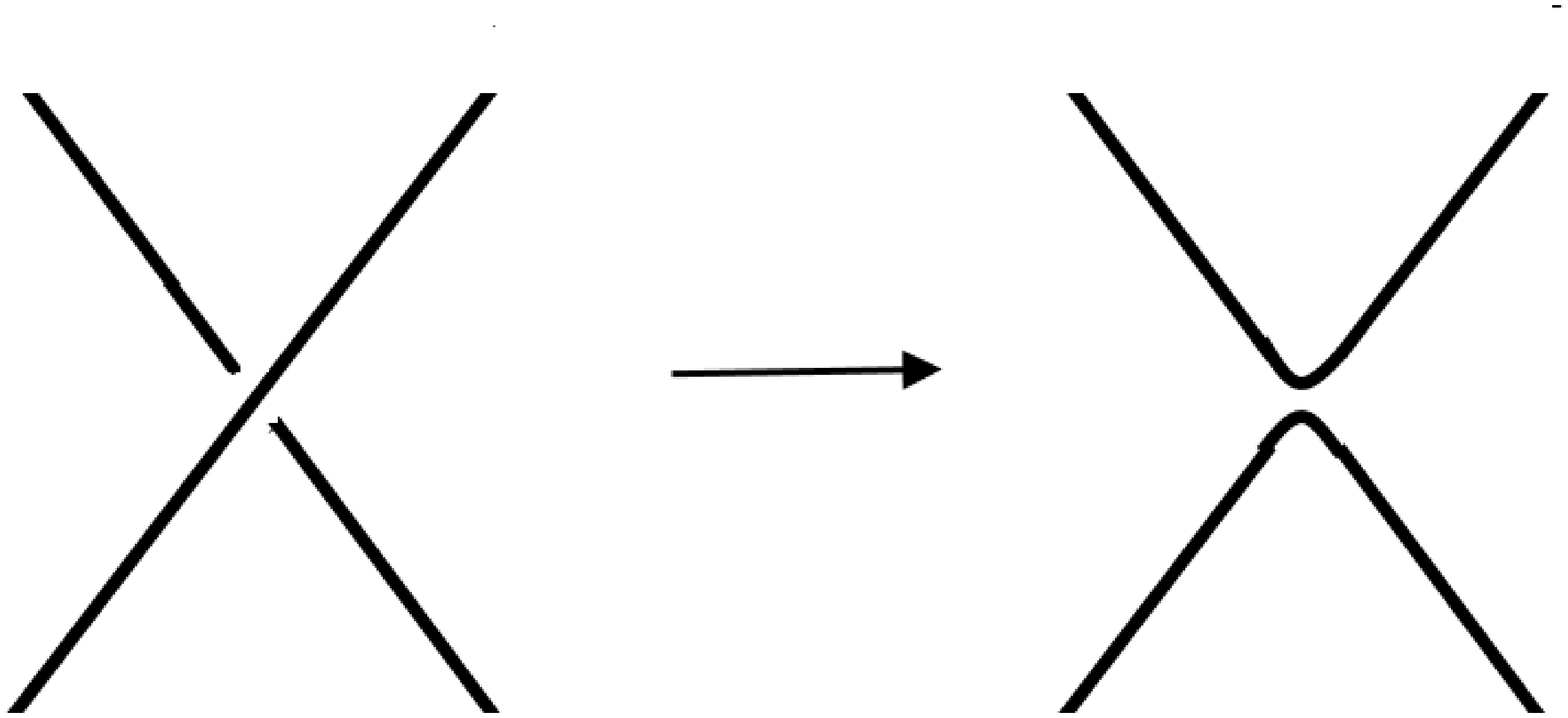}}
\hspace*{20mm}
\subfigure[Loop formation.]{
\includegraphics[width=0.32\textwidth]{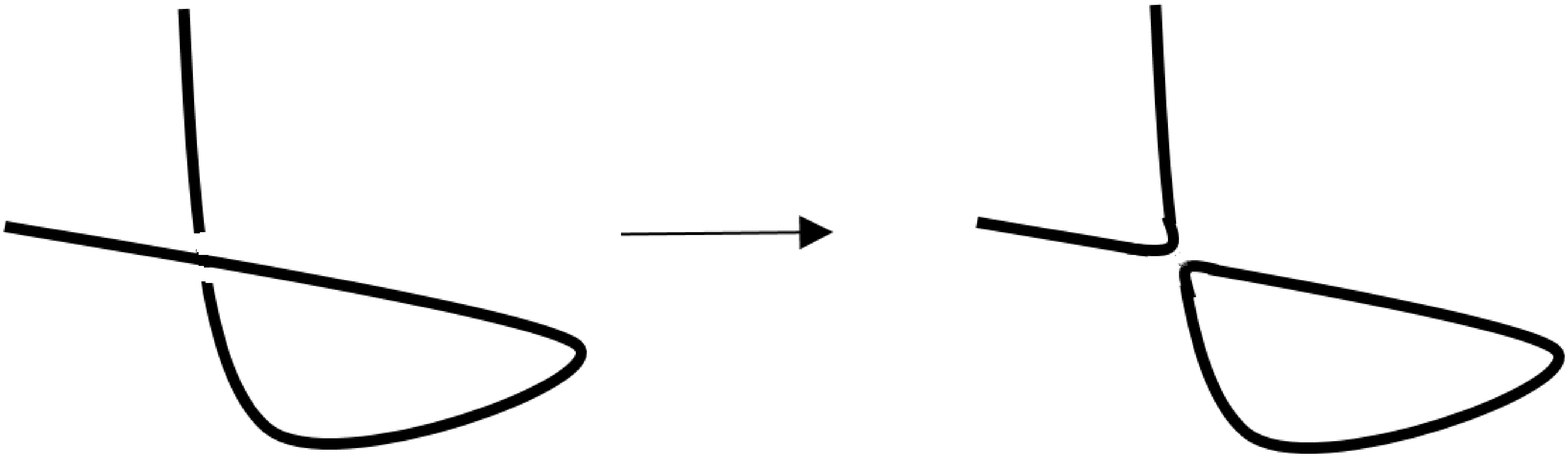}}
\caption{Reconnection processes of cosmic strings.}
\label{Reconnection}
\end{figure}

Major reconnection processes include intercommunication and loop formation, as shown in Fig.\ref{Reconnection}. Tiny twisted circle-shaped objects are produced during the processes, causing topological conservation breaking and energy dissipation via gravitational radiation \cite{Achucarro:2010ub}. After a series of reconnection events, knotted cosmic strings tend to reach a stable stage, where only the strings with much less tangledness are left, and trivial circles and long open strings terminate at the boundary. 

Figs.\ref{crossing} and \ref{Deomposition} above provide us a hint for studying these topologically nonconservative events with energy dissipation. An important fact observed is that, when the crossings
\PICunorientpluscross ~and \PICunorientminuscross ~degenerate to \PICunorientLRsplit ~and \PICunorientUDsplit , they are always accompanied by the emergence of the writhing structures $\gamma_{\pm}$, \PICclosegammaplus ~and \PICclosegammaminus; the same thing happens to the degeneration from the nontrivial \PICopengammaplus ~and \PICopengammaminus ~to the trivial \PIClinecircle. The tiny structures $\gamma_{\pm}$ might play the role of a \textit{topological-conservation breaker} and \textit{lost-energy carrier}, which has potential significance for future study of energy dissipation in reconnection processes of tangled cosmic strings.


\section{Acknowledgements}

We are grateful to Professor Renzo L. Ricca for helpful discussions. X.L. was financially supported by the National Natural Science Foundation of China (NSFC, No.11572005) and the Youth Fellowship of the Haiju Project of Beijing. Y.C.H. was supported by the NSFC (No. 11275017 and 11173028).


\end{document}